\documentclass[12pt,a4paper]{article}
\usepackage{amsfonts,latexsym}
\usepackage{amsmath,amssymb}
\usepackage{graphicx,color}
\usepackage{multirow}
\numberwithin{equation}{section} \oddsidemargin 0 mm \evensidemargin
0 mm \topmargin -10 mm \textheight 215 mm \textwidth 163 mm

\renewcommand{\thefootnote}{\fnsymbol{footnote}}
\newcommand{\nn}{\nonumber}

\begin{document}
\vspace{12mm}

\begin{center}
{{{\Large {\bf Tricritical gravity waves in the four-dimensional generalized massive gravity}}}}\\[10mm]

{Taeyoon Moon$^{a}$\footnote{e-mail address: tymoon@sogang.ac.kr}
and  Yun Soo Myung$^{b}$\footnote{e-mail address:
ysmyung@inje.ac.kr},
}\\[8mm]

{{${}^{a}$ Center for Quantum Space-time, Sogang University, Seoul, 121-742, Korea\\[0pt]
${}^{b}$ Institute of Basic Sciences and School of Computer Aided Science, Inje University Gimhae 621-749, Korea}\\[0pt]
}
\end{center}
\vspace{2mm}

\begin{abstract}
We construct a generalized massive gravity by combining quadratic
curvature gravity with the Chern-Simons  term in four dimensions.
This may be  a candidate for the parity-odd tricritical gravity
theory.  Considering the AdS$_4$ vacuum solution, we derive the
linearized Einstein equation, which is not similar to that of the
three dimensional (3D) generalized massive gravity. When  a
perturbed metric tensor is chosen to be the Kerr-Schild form, the
linearized equation reduces to a single massive scalar equation.  At
the tricritical points where two masses are equal to $-1$ and 2, we
obtain a log-square wave solution to the massive scalar equation.
This is compared  to the 3D tricritical generalized massive gravity
whose dual  is a rank-3 logarithmic conformal field theory.
\end{abstract}
\vspace{5mm}

{\footnotesize ~~~~PACS numbers: }

\vspace{1.5cm}

\hspace{11.5cm}{Typeset Using \LaTeX}
\newpage
\renewcommand{\thefootnote}{\arabic{footnote}}
\setcounter{footnote}{0}


\section{Introduction}

 Stelle~\cite{Stelle} has first introduced the quadratic curvature gravity  of
$a(R_{\mu\nu}^2-R^2/3)+bR^2$ in addition to the Einstein-Hilbert
action $R$  to improve  the perturbative properties of Einstein
gravity.     If $ab\not=0$, the renormalizability was achieved but
the unitarity was violated for  $a\not=0$, indicating that the
renormalizability and unitarity exclude to each other. Although the
$a$-term of providing  massive graviton improves  the ultraviolet
divergence, it induces ghost excitations which spoil  the unitarity
simultaneously.  The price one has to pay for making the theory
renormalizable in this way is the loss of unitarity.  After that
work, the search for a consistent quantum gravity has mainly been
suffered from obtaining a renormalizable and unitary quantum field
theory.

Hence,  a first requirement  for the quantum gravity is to gain the
unitarity, which means that the classical linearized theory has no
tachyon and ghost in the particle content. To that end, critical
gravities have  received  much attention because they were
considered as toy models for quantum
gravity~\cite{Li:2008dq,Lu:2011zk,Deser:2011xc,Porrati:2011ku,
Alishahiha:2011yb,Bergshoeff:2011ri,Lu:2011mw}.  At the critical
point, a degeneracy takes place in the AdS spacetimes and thus,
ghost-like massive gravitons become massless gravitons. In this
case, an equal amount of logarithmic modes should be considered for
the critical gravity instead of massive gravitons. However, one has
to resolve  the non-unitarity problem of the log-gravity
theories~\cite{Lu:2011zk}.
 According to the
AdS/CFT correspondence, it shows that a rank-2 logarithmic conformal
field theory (LCFT) is dual to a critical
gravity~\cite{Grumiller:2008qz,Myung:2008dm,Maloney:2009ck}. Thus,
the non-unitarity of log-gravity is closely related to the
non-unitarity of the LCFT  which arises from the fact that the
Hamiltonian cannot be diagonalized on the fields because of the
Jordan structure~\cite{Gurarie:1993xq,Flohr:2001zs}. In order to
avoid the non-unitarity, one has to truncate out the log-modes by
imposing the  AdS boundary conditions. After truncation, there
remains nothing for the unitary theory.

A polycritical gravity was, recently,  introduced to provide
multiple critical points~\cite{Nutma:2012ss} which might be
described by  a higher rank LCFT.  A rank of the LCFT refers to the
dimensionality of the Jordan cell. For example, the LCFT dual to a
critical gravity has  a rank-2 Jordan cell and thus, an operator has
a logarithmic partner.  The LCFT dual to a tricritical gravity has
rank-3 Jordan cell and  an operator has two logarithmic partners. In
particular, a truncation allows an odd rank LCFT to be a unitary
conformal field theory (CFT)~\cite{Bergshoeff:2012sc}. A 3D
six-derivative tricritical gravity  was treated as a dual to a
rank-3 LCFT~\cite{Bergshoeff:2012ev}, while a four-derivative
critical gravity in four dimensions was considered as a dual to a
rank-3 LCFT~\cite{Johansson:2012fs}.  Furthermore, it was shown that
a consistent truncation of polycritical gravity may be realized at
the linearized level for odd rank~\cite{Kleinschmidt:2012rs}.  On
the other hand, a non-linear tricritical gravity  with  rank-3  was
investigated in three and four dimensions~\cite{Apolo:2012vv}. It is
worth mentioning that a tricritical gravity was first considered  in
a six-derivative gravity in six dimensions~\cite{LPP}.

Interestingly,  a rank-3 and parity odd tricritical theory  was
considered in the context of four-derivative gravity known as
three-dimensional generalized massive gravity (3DGMG)~\cite{LS}. The
3DGMG is obtained by combining   topologically massive gravity
(TMG)~\cite{Deser:1981wh} with  new massive gravity
(NMG)~\cite{Bergshoeff:2009hq}.   Two tricritical points emerged in
the 3DGMG parameter space,  whose dual theory is a rank-3
LCFT~\cite{Grumiller:2010tj}. A truncation of the tricritical 3DGMG
could be  made by  imposing  $Q_L=0$ with $Q_L$ the
Abbott-Deser-Tekin charge for  log-square boundary
conditions~\cite{Bergshoeff:2012ev}. After truncation, a left-moving
sector of CFT remains as a  unitary theory.

At this stage, we would like to point out two related works on the
3DGMG. A log-square mode was found when choosing the Kerr-Schild
metric on the AdS$_3$ background~\cite{AyonBeato:2009yq}, where its
dual LCFT may not be  properly defined. Around the BTZ black hole
background, the authors~\cite{KMP} have confirmed the AdS/LCFT
correspondence existing between the tricritical 3DGMG and a rank-3
finite temperature LCFT  by computing quasinormal frequencies  of
graviton approximately.  However, the 3DGMG is regarded as a toy
model for  tricritical gravities.   Hence, it is desirable  to
obtain a four-dimensional generalized massive gravity (4DGMG).

 In this work, we wish to construct the 4DGMG by combining quadratic curvature gravity with
the Chern-Simons term. Considering the AdS$_4$ vacuum solution
together with the transverse-traceless gauge, we derive a linearized
Einstein equation, which is not a compact form as found in the
3DGMG. Fortunately, taking the Kerr-Schild form for the metric
perturbation, the linearized tensor equation reduces to a
 massive scalar equation on the AdS$_4$. At the tricritical
points, we  obtain  a log-square wave solution which is similar to
the solution found in the 3DGMG~\cite{AyonBeato:2009yq}.  This is
compared to the 3D tricritical GMG whose dual field theory is a
rank-3 LCFT. Furthermore, we compute the linear excitation energy
and conserved charges which turn out to be zero for  log-square wave
solutions.

\section{General massive gravity in four dimensions}
Before we proceed, we remind the reader that  the 3DGMG is obtained
by combining  NMG with  TMG in three dimensions.  Inspired by it, we
introduce the quadratic curvature gravity together with the
Chern-Simons  term \cite{Jackiw:2003pm,Alexander:2009tp} in four
dimensions as
\begin{eqnarray}
S_{\rm 4DGMG}=\frac{1}{16\pi G}\int d^4 x\sqrt{-g}
\Big[R-2\Lambda+\alpha R^2+\beta
R^{cd}R_{cd}+\frac{\theta}{4}{}^{*}RR \Big]\label{Action}
\end{eqnarray}
where $\Lambda$ is a cosmological constant, $\alpha$ and  $ \beta$
are  parameters, while  $\theta$ is a nondynamical field\footnote{We
note that in the dynamical Chern-Simons (DCS) gravity
\cite{Alexander:2009tp,Cardoso:2009pk,Moon:2011fw}, the scalar
$\theta$ is treated as a dynamical field by adding its kinetic and
potential terms.}. The inclusion of $\theta$ is evident from the
fact that the Pontryagin density
${}^{*}RR={}^{*}R^{a~cd}_{~b}R^{b}_{~acd}$  with $
{}^{*}R^{a~cd}_{~b}=\frac{1}{2}\epsilon^{cdef}R^{a}_{~bef}$  is a
total divergence \cite{Jackiw:2003pm,Alexander:2009tp}. The
$\theta$-term corresponds to a parity-violating term because it is
parity odd under the parity operation:
$\hat{P}[{}^{*}RR]=-[{}^{*}RR]$.  We note that $\theta$ should be
included  as $\nabla_c \theta$ in the equation of motion, indicating
that it may spoil the diffeomorphism symmetry [see Eq.(\ref{bian})].

 Varying for $g_{ab}$ on the action (\ref{Action}) leads to
the Einstein equation
\begin{eqnarray} \label{equa1}
R_{ab}-\frac{1}{2}g_{ab}R+\Lambda g_{ab}+E_{ab}+C_{ab}=0,
\end{eqnarray}
where $E_{ab}$ and the 4D Cotton tensor $C_{ab}$  are given
by\footnote{One can find the same expression (\ref{cotton}) in the
literatures \cite{Jackiw:2003pm,Alexander:2009tp}}
\begin{eqnarray}
&&\hspace*{-2em}E_{ab}=\alpha\left(2RR_{ab}-\frac{1}{2}g_{ab}R^2+2g_{ab}\nabla^2
R-2\nabla_{a}\nabla_{b}R\right)\nonumber\\
&& +\beta\left(\frac{1}{2}g_{ab}\nabla^2
R-\nabla_{a}\nabla_{b}R+\nabla^2 R_{ab}+2R_{acbd}R^{cd}
-\frac{1}{2}g_{ab}R_{cd}R^{cd}\right),\label{eab}\\
&&\hspace*{-2em}C_{ab}=\Big(\nabla_{c}~\theta~\epsilon^{cde}_{~~(a}
\nabla_{|e|}R_{b)d}+\frac{1}{2}\nabla_c\nabla_d
~\theta~\epsilon_{(b}^{~~cef}R^{d}_{~~a)ef}\Big).\label{cotton}
\end{eqnarray}
Here, $C_{ab}$ is a traceless and symmetric tensor.
 It is worth noting  that applying $\nabla_a$ to (\ref{equa1}), we
 get a remaining term
 \begin{equation} \label{bian}
 \nabla^aC_{ab}=-\Big[\frac{\nabla_b\theta}{8}\Big]~{}^{*}RR,
 \end{equation}
which should be zero because the Bianchi identity should be
satisfied as the consistency condition.  If not, a breaking of
diffeomorphism-symmetry  is being realized at the level of the
equation of motion.  For  consistency, solutions of this model
should satisfy  $^{*}RR=0$ for $\nabla_b\theta\not=0$ since the
other case of $\nabla_b\theta=0 (C_{ab}=0)$ reduces to the quadratic
curvature gravity. In this case, the diffeomorphism-symmetry  could
be restored dynamically. Explicitly, it  could be achieved in the
equation of motion by  choosing the AdS$_4$-vacuum, even though the
symmetry-breaking may occur at the action level. We note that the
the same  happens when $\theta$ is treated  as a Lagrange
multiplier.

It is easily shown that Eq.(\ref{equa1}) has an AdS$_4$ solution
\begin{equation}
ds^2_{\rm AdS}=\bar{g}_{ab}dx^a
dx^b=-\Big(1+\frac{r^2}{\ell^2}\Big)dt^2+\frac{dr^2}{1+\frac{r^2}{\ell^2}}+r^2d\Omega^2_2
\end{equation}
whose geometrical quantities are given by
\begin{eqnarray}
\bar{R}_{abcd}=\frac{\Lambda}{3}(\bar{g}_{ac}\bar{g}_{bd}
-\bar{g}_{ad}\bar{g}_{bc}),~~~\bar{R}_{ab}=\Lambda
\bar{g}_{ab},~~~\bar{R}=4\Lambda,~~~\Lambda=-\frac{3}{\ell^2}.
\end{eqnarray}
In order to obtain the linearized Einstein  equation, we introduce
the perturbation around the AdS$_4$ background as
\begin{eqnarray} \label{m-p}
g_{ab}=\bar{g}_{ab}+h_{ab}.
\end{eqnarray}
In this case, the linearized equation to (\ref{equa1}) can be
written as
\begin{eqnarray}\label{pert0}
\delta R_{ab}(h)-\frac{1}{2}g_{ab}\delta R(h)-\Lambda h_{ab}+\delta
E_{ab}+\delta C_{ab}(h)=0,
\end{eqnarray}
where the linearized quantities  $\delta R_{ab}(h),~\delta
R(h),~\delta E_{ab}(h)$ \cite{Lu:2011zk} and $\delta C_{ab}(h)$
\cite{MM} are given by
\begin{eqnarray}\label{cottonp0}
\delta R_{ab}(h)&=&\frac{1}{2}\left(\bar{\nabla}^{c}\bar{\nabla}_a
h_{bc}+\bar{\nabla}^{c}\bar{\nabla}_b
h_{ac}-\bar{\nabla}^2h_{ab}-\bar{\nabla}_a \bar{\nabla}_b h\right), \nn \\
\delta
R(h)&=&\bar{\nabla}^{a}\bar{\nabla}^{b}h_{ab}-\bar{\nabla}^2h-\Lambda
h,\nn\\
\delta E_{ab}(h)&=&
4\Lambda\left(2\alpha+\frac{1}{3}\beta\right)\delta
G_{ab}+(2\alpha+\beta)\left(g_{ab}\bar{\nabla}^2-\bar\nabla_{a}
\bar\nabla_{b}+\Lambda
g_{ab}\right)\delta R\nn\\
&&+\beta\left(\bar{\nabla}^2\delta
G_{ab}-\frac{2\Lambda}{3}g_{ab}\delta R\right),
\nn\\
\delta
C_{ab}(h)&=&\Bigg[\frac{1}{2}v_{c}~\epsilon^{cde}_{~~~a}\left(\bar{\nabla}_{e}\delta
R_{bd}-\Lambda \bar{\nabla}_{e}h_{bd}\right)+\frac{1}{8}v_{cd}~
\epsilon_{b}^{~cef}\Big(\bar{\nabla}_e\bar{\nabla}_f
h^{d}_{~a}+\bar{\nabla}_e\bar{\nabla}_a h^{d}_{~f}\nn\\
&&-\bar{\nabla}_e\bar{\nabla}^d h_{af}-\bar{\nabla}_f\bar{\nabla}_e
h^{d}_{~a}-\bar{\nabla}_f\bar{\nabla}_a
h^{d}_{~e}+\bar{\nabla}_f\bar{\nabla}^d
h_{ae}\Big)\Bigg]+\Bigg[a\leftrightarrow b\Bigg]
\end{eqnarray}
with \begin{equation} \label{leinstein}\delta G_{ab}=\delta
R_{ab}-\frac{1}{2}\delta R \bar{g}_{ab}-\Lambda
h_{ab},~~v_c=\bar{\nabla}_c
\theta,~~v_{cd}=\bar{\nabla}_c\bar{\nabla}_d \theta.\end{equation}
Let us choose the transverse gauge as
\begin{eqnarray}\label{gauge}
\bar{\nabla}^{a}h_{ab}=\bar{\nabla}_bh.
\end{eqnarray}
Substituting (\ref{gauge}) into Eq.(\ref{pert0}), the trace  of
 the linearized equation (\ref{pert0}) becomes
\begin{eqnarray}\label{pert}
\Lambda\Big[h-2(3\alpha+\beta)\bar\nabla^2h\Big]=0
\end{eqnarray}
when considering the traceless condition of $\delta C^a~_a=0$.  In
order to eliminate a massive spin-0 mode, we choose \cite{Lu:2011zk}
\begin{eqnarray}
3\alpha+\beta=0.\label{ab}
\end{eqnarray}
 In this case, Eq.(\ref{pert}) yields $h=0$ which implies that
one  may choose  the transverse traceless (TT) gauge
\begin{eqnarray}\label{gauge0}
\bar{\nabla}^{a}h_{ab}=0,~~~h=0.
\end{eqnarray}
Substituting   (\ref{ab}) and (\ref{gauge0}) into (\ref{pert0})
leads to
\begin{eqnarray}\label{mainp}
\frac{3\alpha}{2}\left(\bar\nabla^2-\frac{2\Lambda}{3}\right)
\left(\bar\nabla^2-\frac{4\Lambda}{3}-\frac{1}{3\alpha}\right)h_{ab}+\delta
C_{ab}(h)=0.
\end{eqnarray}
 Here the linearized Cotton  tensor
$\delta C_{ab}(h)$ takes the complicated form
\begin{eqnarray}\label{cottonp}
\delta
C_{ab}(h)&=&\Big[-\frac{1}{4}v_{c}~\epsilon^{cde}_{~~~a}\bar{\nabla}_e\bar{\nabla}^2h_{bd}+
\frac{\Lambda}{6}v_c~ \epsilon^{cde}_{~~~a}\bar{\nabla}_e
h_{bd}+\frac{1}{4}v_{cd}~\epsilon_{b}^{~cef}\Big(\bar{\nabla}_e\bar{\nabla}_a
h^{d}_{~f}-\bar{\nabla}_e\bar{\nabla}^d h_{af}\Big)\Big]\nn\\
&+&\Big[a\leftrightarrow b\Big].
\end{eqnarray}
At this stage, we  point out that it is difficult  to combine
$\delta C_{ab}$ with  the first term in (\ref{mainp}) without
choosing a specific form of  $\theta$.   It is  known that for a
particular $\theta$~\cite{MM,beato},  one can manipulate $\delta
C_{ab}$ so that the linearized equation (\ref{mainp}) can be
rewritten compactly. Actually, it could be achieved by choosing
\begin{eqnarray}\label{ansat}
\bar{\theta}=k\frac{x}{y},~~~ \bar{g}_{ab}\equiv
\phi^{-2}\eta_{ab}=\frac{\ell^2}{y^2}\eta_{ab}
\end{eqnarray}
in the Poincare coordinates $(u,v,x,y)$ for the AdS$_4$. Here $k$
has the dimension of [mass]$^{-2}$, $\ell$ is the AdS$_4$ curvature
radius, and $\eta_{ab}$ is the flat metric tensor defined as
\begin{eqnarray}
\eta_{ab}dx^{a}dx^{b}=2dudv+dx^2+dy^2.
\end{eqnarray}

Plugging (\ref{ansat}) into  (\ref{mainp}),  one arrives at
\begin{eqnarray}\label{Master1}
\Bigg(\bar{\nabla}^2-\frac{2}{3}\Lambda\Bigg)
\Bigg[-3\alpha\Big(\bar{\nabla}^2-\frac{4}{3}\Lambda
-\frac{1}{3\alpha}\Big)h_{ab}
+v_c~\epsilon^{cde}_{~~~(a}\bar{\nabla}_{|e|}h_{b)d}\Bigg]=0.
\end{eqnarray}
Alternatively, it leads to
\begin{eqnarray}\label{Master1-1}
\Bigg[-3\alpha\Big(\bar{\nabla}^2-\frac{4}{3}\Lambda
-\frac{1}{3\alpha}\Big) \delta_{(a}^{a^{'}}\delta_{b)}^{d}
+\delta_{(a}^{a^{'}}v_{|c|}~\epsilon^{cde}_{~~~b)}\bar{\nabla}_e\Bigg]
\Bigg(\bar{\nabla}^2-\frac{2}{3}\Lambda\Bigg)h_{a^{'}d}=0
\end{eqnarray}
because  two operations in Eq.(\ref{Master1}) commute to each other.
Here  $v_c$ is determined  by
\begin{equation}\label{vc}
v_c=k\Bigg(0,0,\frac{1}{y},-\frac{x}{y^2}\Bigg) \end{equation} which
may generate the mass of graviton.  In this case, $v_c$ is not a
constant vector but a vector field.

\section{AdS wave as perturbation}
It is a formidable task to solve (\ref{Master1-1}) directly because
the 4D Cotton term induces a complicated expression upon choosing
$v_c$ in (\ref{vc}).
 In order to solve  (\ref{Master1-1}), we take the Kerr-Schild
 form as an AdS$_4$-wave solution
\begin{eqnarray}\label{metric1}
h_{ab}=2\Phi\lambda_{a}\lambda_{b},
\end{eqnarray}
where $\lambda_{a}=(1,0,0,0)$ is a null ($\lambda^2=0$) and geodesic
vector and a scalar field $\Phi=\Phi(u,v,x,y)$. Additionally, the TT
gauge condition (\ref{gauge0}) implies that one may  restrict $\Phi$
to $\Phi(u,x,y)$ by taking account of the condition of
$\lambda_{a}\bar{\nabla}^{a}\Phi=0$.  Substituting
$h_{ab}=2\Phi\lambda_a\lambda_b$ into  the tensor equation
(\ref{Master1}) leads to the scalar equation for $\Phi$
\begin{eqnarray} \label{adsteneq}
&&\hspace*{-3em}\lambda_{a'}\lambda_{d}\Big[-3\alpha\Big(\bar{\nabla}^2
+\frac{4}{\phi}\partial^{f}\phi ~\partial_{f}-\frac{1}{3\alpha}
\Big)\delta_{(a}^{a^{'}}\delta_{b)}^{d}
+\delta_{(a}^{a^{'}}v_{|c|}~\epsilon^{cde}_{~~~b)}
\Big(\frac{\partial_e\phi}{\phi}+\bar{\nabla}_e\Big)\Big]\nn\\
&&\hspace*{17em}\times\Big[\bar{\nabla}^2
+\frac{2}{3}\Lambda+\frac{4}{\phi}\partial^{f}\phi
~\partial_{f}\Big]\Phi=0.
\end{eqnarray}
 Introducing the separation of variables
$\Phi(u,x,y)=U(u)X(x)Y(y)$ and taking into account $\lambda_a,~\phi$
in (\ref{ansat}) and $v_c$ in (\ref{vc}), Eq.(\ref{adsteneq})
reduces to
\begin{eqnarray}\label{master3}
&&\hspace*{-3em}\Bigg[3\alpha\Big(y^2(\partial_y^2+\partial_x^2)
+2y\partial_y-\frac{\ell^2}{3\alpha}\Big)
+k\Big(x\partial_x+y\partial_y+1\Big)\Bigg]\\
\label{master31}&&\hspace*{13em}\times\Bigg(y^2(\partial_y^2+\partial_x^2)
+2y\partial_y-2\Bigg)XY=0.
\end{eqnarray}

On the other hand,  as was suggested in the 3DGMG~\cite{LS},
(\ref{master3})  includes  two mass parameters $m_1$ and $m_2$
implicitly.  In order to obtain two massive equations from
(\ref{master3}),  we introduce the four one-dimensional operators,
\begin{eqnarray}
D^{m_i}=y\partial_{y}+m_i,~~~D^{-1}=y\partial_{y}-1,~~~
D^{+2}=y\partial_{y}+2
\end{eqnarray}
with $i=1,~2$.  Using these mutually commuting operators,
(\ref{master3}) and (\ref{master31}) can be expressed compactly as
\begin{eqnarray} \label{difform}
D^{m_1}D^{m_2}D^{-1}D^{+2}~Y(y)=0 \to
(y\partial_{y}+m_1)(y\partial_{y}+m_2)
(y\partial_{y}-1)(y\partial_{y}+2)~Y(y)=0.\label{dddd}
\end{eqnarray}
In deriving this, we assume  $X(x)=$constant for simplicity.
Comparing Eq.(\ref{dddd}) with Eq.(\ref{master3}), we find $m_1$ and
$m_2$ for $\alpha\neq 0$
\begin{eqnarray}
m_1&=&\frac{1}{2}\Bigg(1+\frac{k}{3\alpha}
+\sqrt{\frac{k^2}{9\alpha^2}-\frac{2k}{3\alpha}+1
+\frac{4\ell^2}{3\alpha}}\Bigg),\\
m_2&=&\frac{1}{2}\Bigg(1+\frac{k}{3\alpha}
-\sqrt{\frac{k^2}{9\alpha^2}-\frac{2k}{3\alpha}+1
+\frac{4\ell^2}{3\alpha}}\Bigg).
\end{eqnarray}
For $\alpha=0$, being the Chern-Simons gravity~\cite{MM}, we have a
third-order linearized  equation from Eq.(\ref{master3}) instead of
the fourth-order equation (\ref{dddd}).  In this case, a single mass
parameter $m$ is given by either
\begin{eqnarray}\label{pt45}
m_1=1-\frac{\ell^2}{k}~~~{\rm or}~~~m_2=1-\frac{\ell^2}{k}.
\end{eqnarray}

\section{Tricritical and critical points}

Now we wish to obtain  tricritical points.  For
 this purpose,  we   focus on two points
\begin{eqnarray}
{\rm point~ 1}~:&&k=9\alpha=-3\ell^2~\Rightarrow~m_1=m_2=2\\
{\rm point~
2}~:&&k=-9\alpha=\frac{3}{4}\ell^2~\Rightarrow~m_1=m_2=-1.
\end{eqnarray}
For each point, (\ref{dddd}) reduces to
\begin{eqnarray}
D^{-1}D^{+2}D^{+2}D^{+2}~Y(y)=0
&\Rightarrow&(y\partial_{y}+2)^3(y\partial_{y}-1)~Y(y)=0\label{d2}\\
D^{-1}D^{-1}D^{-1}D^{+2}~Y(y)=0
&\Rightarrow&(y\partial_{y}-1)^3(y\partial_{y}+2)~Y(y)=0,
\label{d-1}
\end{eqnarray}
respectively.  It is important to note that at the point 1,
$D^{m_1}$ and $D^{m_2}$ are  $D^{+2}$ [Eq.(\ref{d2})], while at the
point 2, they are  $D^{-1}$ [Eq.(\ref{d-1})]. We remind the reader
that these are  two tricritical points, implying  that there exists
a three-fold degeneracy.

On the other hand,  the quadratic curvature
gravity~\cite{Gullu:2011sj}  obtained from taking the $k=0$ limit in
Eq.(\ref{master3}) has  a critical point of
$\alpha=\ell^2/6=-1/(2\Lambda)$.  In this case, (\ref{master3}) and
(\ref{master31}) reduce to
\begin{eqnarray}
D^{-1}D^{+2}D^{-1}D^{+2}~Y(y)=0
~\Rightarrow~(y\partial_{y}-1)^2(y\partial_{y}+2)^2~Y(y)=0.
\end{eqnarray}
In the Chern-Simons  gravity  which is equivalent to the  $\alpha=0$
limit of  (\ref{master3})~\cite{MM}, we obtain two critical points
of $k=\ell^2/2$ and $k=-\ell^2$ from Eq.(\ref{pt45}).  For each $k$,
 (\ref{master3}) and (\ref{master31})  become the third-order
equation as
\begin{eqnarray}
D^{-1}D^{-1}D^{+2}~Y(y)=0
&\Rightarrow&(y\partial_{y}+2)(y\partial_{y}-1)^2~Y(y)=0,~~~(k=\ell^2/2)\\
D^{-1}D^{+2}D^{+2}~Y(y)=0
&\Rightarrow&(y\partial_{y}-1)(y\partial_{y}+2)^2~Y(y)=0.
~~~(k=-\ell^2)\label{cs2}
\end{eqnarray}
For $k=\ell^2/2$, a two-fold degeneracy emerges at $m=m_1=-1$,
whereas there exists the other degeneracy at $m=m_2=2$ for
$k=-\ell^2$.
\begin{figure*}[t!]
   \centering
   \includegraphics{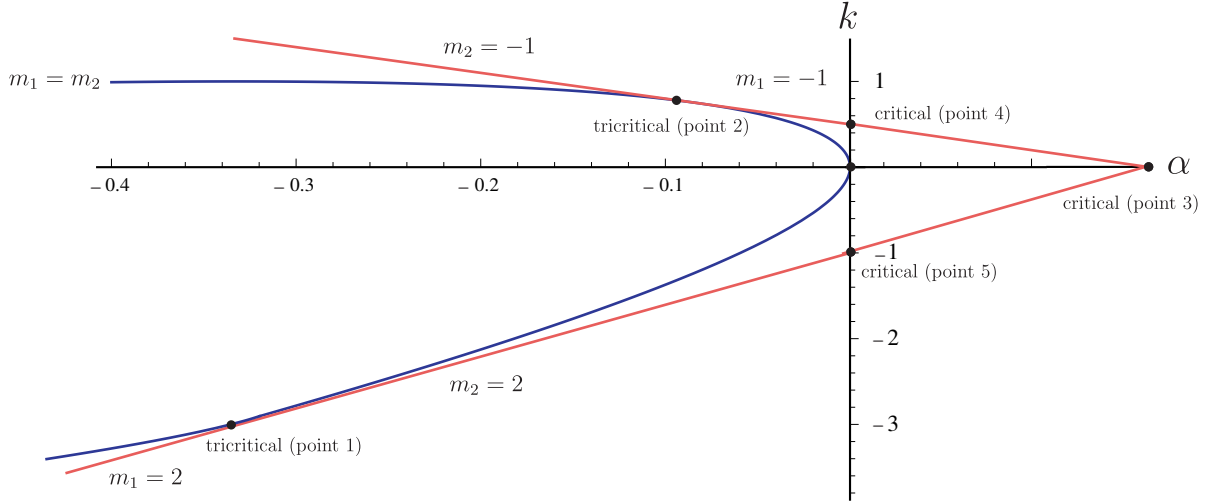}
\caption{The tricritical and critical points in the parameter space
($\alpha,k$)   for $\ell^2=1$. The tricritical points $1$ and $2$
correspond to $(\alpha,~k)=(-1/3,~-3)$ and $(-1/12,~3/4)$,
respectively,  while three critical points 3, 4, 5 correspond  to
($1/6,~0$), ($0,~1/2$), and ($0,~-1$), respectively. In addition,
the origin (0,~0) denotes the Einstein gravity.} \label{CS-2}
\end{figure*}

 Fig. 1 shows two tricritical points and three critical
points in the parameter space $(\alpha,~k)$ of the 4DGMG.  We find
the solutions to Eqs.(\ref{d2})-(\ref{cs2})
for each point:\\
\\
(i) $m_1=m_2=2$  $({\rm point~1:tricritical})$
\begin{eqnarray}
\Phi(u,y)&=&U(u)Y(y)\nn\\
&=&c_1(u)y+\frac{1}{y^2}\Big[c_2(u)+c_3(u)\ln(y)
+c_4(u)\ln^2(y)\Big], \label{logs1}
\end{eqnarray}
\\
(ii) $m_1=m_2=-1$  $({\rm point~ 2:tricritical})$\\
\begin{eqnarray}
\Phi(u,y)&=&U(u)Y(y)\nn\\
&=&c_5(u)\frac{1}{y^2}+y\Big[c_6(u)+c_7(u)\ln(y)
+c_8(u)\ln^2(y)\Big], \label{logs2}
\end{eqnarray}
\\
(iii) $m_1=2$, $m_2=-1$ $({\rm point~ 3:critical})$
\begin{eqnarray}
\Phi(u,y)&=&U(u)Y(y)\nn\\
&=&d_1(u)y+d_2(u)\frac{1}{y^2}+d_3(u)y\ln(y)
+d_4(u)\frac{\ln(y)}{y^2},\label{sol33}
\end{eqnarray}
\\
(iv) $m_1=-1$, $m_2=0$ $({\rm point~4:~critical})$
\begin{eqnarray}
\Phi(u,y)&=&U(u)Y(y)\nn\\
&=&d_5(u)y+d_6(u)\frac{1}{y^2}+d_7(u)y\ln(y),\label{sol44}
\end{eqnarray}
\\
(v) $m_1=0$, $m_2=2$ $({\rm point~5:~critical})$
\begin{eqnarray}
\Phi(u,y)&=&U(u)Y(y)\nn\\
&=&d_8(u)y+d_9(u)\frac{1}{y^2}+d_{10}(u)\frac{\ln(y)}{y^2},\label{sol55}
\end{eqnarray}
where all of $c_{i}$ and $d_{i}$ are undetermined functions of $u$.

From (i)-(v), we observe  that at  tricritical points $1$ and $2$,
there exist log-square  terms, as was shown in the tricritical
3DGMG~\cite{AyonBeato:2009yq}, while at the critical points  3, 4,
5, the solutions have  log-terms. In particular,  combination of
(\ref{sol44}) with  (\ref{sol55}) yields the solution (\ref{sol33})
found in the quadratic curvature gravity~\cite{Gullu:2011sj} which
provides the fourth-order perturbation equation at the critical
point.

\section{Linearized excitation energy and conserved charges}
In this section, we compute  the linearized excitation energy of the
log-square wave solution (\ref{metric1}) with (\ref{logs1}) and
(\ref{logs2}) to check whether the ghost state exists or not. Also,
 we find the conserved charges. In order to calculate the excitation energy, we first
construct the bilinear Hamiltonian $H^{(2)}$. The bilinear action
for the metric perturbation  around the AdS$_4$ spacetimes is given
by
\begin{eqnarray}\label{2action}
S^{(2)}&=&-\frac{1}{16\pi G}\int
d^4x\sqrt{-\bar{g}}h^{ab}\Big[\delta
R_{ab}-\frac{1}{2}\bar{g}_{ab}\delta R-\Lambda
h_{ab}+\delta E_{ab}+\delta C_{ab}\Big]\nn\\
&=&-\frac{1}{16\pi G}\int
d^4x\sqrt{-\bar{g}}\Big[\frac{3}{2}\alpha(\bar{\nabla}^2h^{ab})
(\bar{\nabla}^2h_{ab})+\left(\frac{1}{2}+3\Lambda\alpha\right)
(\bar{\nabla}^{c}h^{ab})(\bar{\nabla}_{c}h_{ab})
\nn\\
&&\hspace*{-3em}+\frac{\Lambda}{3}\left(4\Lambda\alpha+1\right)h^{ab}h_{ab}
+\frac{1}{2}\epsilon^{cde}_{~~~a}
\Big(v_{ce}h^{ab}\bar{\nabla}^2h_{bd}
+v_c\bar{\nabla}_eh^{ab}\bar{\nabla}^2h_{bd} +\frac{2}{3}\Lambda v_c
h^{ab}\bar{\nabla}_eh_{bd}\Big)\nn\\
&&\hspace*{-3em}-\frac{1}{2}\epsilon_b^{~cef}\Big(v_{cde}
h^{ab}\bar{\nabla}_{a}h^{d}_{~f}+v_{cd}\bar{\nabla}_{e}h^{ab}\bar{\nabla}_{a}h^{d}_{~f}
-v_{cde}h^{ab}\bar{\nabla}^{d}h_{af}-v_{cd}\bar{\nabla}_{e}h^{ab}\bar{\nabla}^{d}h_{af}\Big)\Big].
\end{eqnarray}
In deriving (\ref{2action}), we have used the transverse-traceless
gauge (\ref{gauge0}).  From the above bilinear action, the canonical
momentum is  defined by
\begin{eqnarray}
\Pi^{ab}_{(1)}&=&\frac{\delta{\cal L}^{(2)}}{\delta\dot{h}_{ab}}
-\bar{\nabla}_0\left(\frac{\delta{\cal L}^{(2)}}{\delta(d(\bar{\nabla}_0h_{ab})/dt)}\right)\nn\\
&=&-\frac{1}{16\pi
G}\sqrt{-\bar{g}}\Bigg[\bar{\nabla}^{0}(1+6\Lambda\alpha)h^{ab}
-3\alpha\bar{g}^{00}\bar{\nabla}_{0}\bar{\nabla}^2h^{ab}+\frac{1}{2}\epsilon^{cd0a}v_c
\bar{\nabla}^2h^{b}_{~d}+\frac{\Lambda}{3}v_c\epsilon^{ca0}_{~~~d}h^{db}
\nn\\
&&-\frac{1}{2}\epsilon^{bc0f}v_{cd}\bar{\nabla}^{a}h^{d}_{~f}
-\frac{1}{2}\epsilon_{f}^{~ceb}v_c^{a} \bar{\nabla}_e
h^{0f}+\frac{1}{2}\epsilon_f^{~ceb}v_{c~e}^{~0}h^{af}
+\frac{1}{2}\epsilon^{bc0f}v_{cd}\bar{\nabla}^{d}h^{a}_{~f}
\nn\\
&&+\frac{1}{2}\epsilon_f^{~ceb}v_c^{0}
\bar{\nabla}_{e}h^{af}-\frac{1}{2}\bar{\nabla}_{0}
(\epsilon^{cae}_{~~~d}v_c\bar{\nabla}_{e}h^{db}\bar{g}^{00})\Bigg]
\end{eqnarray}
with $S^{(2)}=\int d^4x {\cal L}^{(2)}$. Using the Ostrogradsky
formalism for a higher-order derivative Lagrangian, one defines the
canonical momentum for the second-order temporal derivative as
\begin{eqnarray}
\Pi^{ab} _{(2)}&=&\frac{\delta{\cal
L}^{(2)}}{\delta(d(\bar{\nabla}_0h_{ab})/dt)}
\nn\\&=&-\frac{1}{32\pi
G}\sqrt{-\bar{g}}\bar{g}^{00}\Big(6\alpha\bar{\nabla}^2h^{ab}
+\epsilon^{cae}_{~~~d}v_c\bar{\nabla}_{e}h^{db}\Big).
\end{eqnarray}
Then, the bilinear Hamiltonian takes the form
\begin{eqnarray}\label{hamil}
H^{(2)}=\int
d^{4}x\Big(\dot{h}_{ab}\Pi^{ab}_{(1)}+\frac{\partial}{\partial
t}(\bar{\nabla}_0h_{ab})\Pi^{ab}_{(2)}\Big)-S^{(2)}.
\end{eqnarray}
 Substituting Eqs.(\ref{ansat}) and  (\ref{metric1})
 into Eq.(\ref{hamil}) and   using
 $\lambda^2=0$ and $\lambda^{a}\bar{\nabla_a}\lambda_b=0$ leads to
 the vanishing Hamiltonian
 \begin{equation}
H^{(2)}=0, \end{equation} which shows  that
 there is no ghost states for the log-square waves (\ref{logs1}) and (\ref{logs2}).

On the other hand,  one may use  the covariant formalism for 4D
Chern-Simons gravity~\cite{Tekin:2007rn} and quadratic curvature
gravity~\cite{Deser:2011xc} in asymptotically AdS$_4$ spacetimes to
derive  the conserved charges in the 4DGMG as
\begin{eqnarray}\label{charge}
Q^{a}=\int_{\cal M}dS_i\sqrt{-\bar{g}}\Bigg[{\cal A}^{a
i}(\bar{\eta})+(1+2\Lambda\alpha){\cal A}^{a
i}(\bar{\xi})+\alpha{\cal B}^{a i}(\bar{\xi})-\frac{1}{2}{\cal C}^{a
i}(\bar{\xi})\Bigg],
\end{eqnarray}
where $\bar{\eta}^{a} =v_{b}\epsilon^{bcda}
\bar{\nabla}_{d}\bar{\xi}_{c}/2$ with a Killing vector
$\bar{\xi}_{c}$
 and
\begin{eqnarray}
{\cal A}^{a i}(\bar{\xi})&=&\bar{\xi}_{b}\bar{\nabla}^{a}h^{b
i}-\bar{\xi}_{b}\bar{\nabla}^{i}h^{ab}
+\bar{\xi}^{a}\bar{\nabla}^{i}h-\bar{\xi}^{i}\bar{\nabla}^{a}h
+h^{ab}\bar{\nabla}^{i}\bar{\xi}_{b}-h^{b
i}\bar{\nabla}^{a}\bar{\xi}_{b}
\nn\\
&&+\bar{\xi}^{i}\bar{\nabla}_{b}h^{ab}-\bar{\xi}^{a}\bar{\nabla}_{b}h^{ib}
+h\bar{\nabla}^{a}\bar{\xi}^{i},\nn\\
{\cal B}^{a i}(\bar{\xi})&=&-\bar{\xi}^{a}\bar{\nabla}^{i}\delta
R-\delta
R\bar{\nabla}^{a}\bar{\xi}^{i}+\bar{\xi}^{i}\bar{\nabla}^{a}\delta
R-3\Big(\bar{\xi}_{b}\bar{\nabla}^{i}\delta
G^{ab}-\bar{\xi}_{b}\bar{\nabla}^{a}\delta G^{b
i}\nn\\
&&-\delta G^{ab}\bar{\nabla}^{i}\bar{\xi}_{b}+\delta G^{b
i}\bar{\nabla}^{a}\bar{\xi}_{b}\Big),\nn\\
{\cal C}^{a i}(\bar{\xi})&=&v_{b}\bar{\xi}^{d}\epsilon^{ba ic}\delta
G_{dc}+v_{b}\bar{\xi}_{c}\epsilon^{b c id}\delta
G^{a}_{d}+v_{d}\bar{\xi}_{b}\epsilon^{d a bc}\delta G^{i}_{c}.
\end{eqnarray}
In these expressions, $\delta R$ and $\delta G_{ab}$ are given by
 (\ref{cottonp0}) and (\ref{leinstein}). Plugging (\ref{ansat})
and  (\ref{metric1})
 into (\ref{charge}) together with the Killing vector
 $\bar{\xi}_{a}=(1,0,0,0)/y^2$ and $\bar{\eta}^{a}=k\Lambda^2(0,1,0,0)/9$,
 then four quantities
 ${\cal
A}^{a i}(\bar{\xi}),~{\cal A}^{a i}(\bar{\eta}),~{\cal B}^{a
i}(\bar{\xi}),$ and ${\cal C}^{a i}(\bar{\xi})$ take simpler forms
\begin{eqnarray}
\hspace*{-2em}{\cal A}^{a
i}(\bar{\xi})&=&\frac{2\Phi}{y^2}{\cal M}^{ai},\label{A1}\\
\hspace*{-2em}{\cal A}^{a
i}(\bar{\eta})&=&\frac{2k\Lambda^2\Phi}{9}{\cal M}^{ai},\label{A2}\\
\hspace*{-2em}{\cal B}^{a i}(\bar{\xi})&=&
-\frac{3}{y^2}\Big(\bar{\nabla}^2\Phi-\frac{2}{3}\Lambda\Phi
-\frac{4}{3}\Lambda y \Phi^{\prime}\Big){\cal M}^{ai},\label{B}\\
\hspace*{-2em}{\cal C}^{a i}(\bar{\xi})&=&-\frac{v_{b}}{y^2}
\Big(\bar{\nabla}^2\Phi-\frac{2}{3}\Lambda\Phi -\frac{4}{3}\Lambda y
\Phi^{\prime}\Big) \Big(\lambda^{d}\epsilon^{ba
ic}\lambda_{d}\lambda_{c}+\lambda_{c}\epsilon^{b c id}
\lambda^{a}\lambda_{d}+\lambda_{b}\epsilon^{d a bc}
\lambda^{i}\lambda_{c}\Big).
\end{eqnarray}
Here the prime (${}^{\prime}$) denotes the differentiation with
respect to $y$ and ${\cal M}^{ai}$ is given by the null vector
$\lambda^a$
\begin{eqnarray}
{\cal M}^{ai}=\lambda_{b}\bar{\nabla}^{a}
(\lambda^b\lambda^i)-\lambda_{b}\bar{\nabla}^{i}(\lambda^{a}\lambda^{b})
+\lambda^{a}\lambda^{b}\bar{\nabla}^{i}\lambda_{b}-\lambda^{b}\lambda^{i}
\bar{\nabla}^{a}\lambda_{b}.
\end{eqnarray}
In deriving these, we have used a relation of
\begin{eqnarray}
\bar{\nabla}^2(\Phi\lambda_{a}\lambda_{b})
=\Big(\bar{\nabla}^2\Phi+\frac{4}{3}\Lambda\Phi-\frac{4}{3}\Lambda
y\Phi^{\prime}\Big)\lambda_{a}\lambda_{b}.
\end{eqnarray}
Employing two relations
\begin{eqnarray}
\bar{\nabla}_{a}\lambda_b&=&
\frac{1}{\phi}(\lambda_a\partial_b\phi+\lambda_b\partial_a\phi),
\\
\bar{\nabla}_{c}(\lambda_{a}\lambda_{b})
&=&\frac{1}{\phi}(2\lambda_{a}\lambda_{b}\partial_{c}\phi
+\lambda_{c}\lambda_{b}\partial_{a}\phi
+\lambda_{c}\lambda_{a}\partial_{b}\phi),
\end{eqnarray}
we  obtain ${\cal C}^{a i}(\bar{\xi})=0$ and ${\cal M}^{ai}=0$ which
implies that ${\cal A}^{a i}(\bar{\xi})={\cal A}^{a
i}(\bar{\eta})={\cal B}^{a i}(\bar{\xi})=0$. Finally, we arrive at
the result that the conserved charges vanish
\begin{equation} Q^{a}=0.
\end{equation}

Before closing this section, it is worth explaining  why the
conserved charges  vanish.  It is well-known that the conserved
charges of the energy and angular momentum  can be obtained through
the integration at spatial infinity  if   a timelike and a
rotational Killing vector are properly given.  It suggests that in
our case, the conserved charges of energy and angular momentum
vanish  because the AdS-wave admits a nulllike Killing vector
$\bar{\xi}_c$ only.  Also we would like to mention that the zero
excitation energy and zero charges are  due to the following
sequential restriction from tensor to scalar:
\begin{equation}
h_{ab}~~\xrightarrow[{\rm Kerr-Schild~ form}]{}~~2\Phi
\lambda_a\lambda_b~~ \xrightarrow[{\rm gauge-fixing}]{}~~
\lambda_a\bar{\nabla}^a \Phi=0.
\end{equation}
Hence, we conjecture that all AdS-wave solutions
(\ref{logs1})-(\ref{sol55}) including non-critical solutions do not
provide the nonzero excitation energy and conserved charges.

\section{Discussions}
We have constructed the  generalized massive gravity by combining
quadratic curvature  gravity with the Chern-Simons term in four
dimensions.  This 4DGMG is similar to the 3DGMG. Considering the
AdS$_4$ vacuum solution, we have derived the linearized Einstein
equation, which is not a compact form  compared with that of the
3DGMG. When the metric perturbation is chosen to be the Kerr-Schild
form, however, the linearized tensor equation reduces to a massive
scalar equation.  At the tricritical points, we obtain a log-square
wave solution  whose dual field theory may  not be properly defined.
The absence of its dual LCFT is compared  to the 3D tricritical GMG
whose dual is a rank-3 LCFT.

A difference between the 4DGMG and 3DGMG is that the former has the
Chern-Simons term of  $\frac{\theta}{4}~ ^{*}RR$ with $\theta$ a
nondynamical field,  while the latter has the topologically massive
term $\frac{1}{2\mu}\epsilon \cdot [\Gamma
\partial \Gamma+\frac{2}{3} \Gamma\Gamma\Gamma]$ with $\mu$ a parameter, even though two
terms belong to parity-violating term.   The presence of $\theta$ in
the 4DGMG may prevent us from having its dual LCFT at the
tricritical point because it is not a constant like $1/\mu$ in the
3DGMG but a nondynamical field.  Most of all, choosing $\theta=k
x/y$ makes the linearized equation complicated on the AdS$_4$
spacetimes,  which  implies that the linearized  equation
(\ref{mainp}) cannot be written by a compact form like as  in the
AdS$_3$ spacetimes
\begin{eqnarray}
({\cal D}^L{\cal D}^R{\cal D}^{M_+}{\cal D}^{M_-}h)_{\mu\nu}=0.
\end{eqnarray}
Here $h_{\mu\nu}$ is the metric perturbation  around the AdS$_3$ and
 operators (${\cal D}^{L/R},{\cal D}^{M_\pm}$) are given
in \cite{Bergshoeff:2012ev,LS,KMP}. On the other hand,  the compact
equation (\ref{difform}) expressed in one-dimensional operators is
obtained  only after  choosing the AdS-wave.  This implies that the
parity-odd tricritical gravity theory in the 4DGMG could not be
described by  its dual higher rank LCFT.

 However,  it was
discussed  in the 4D quadratic  curvature
gravity~\cite{Alishahiha:2011yb} that at the critical point, the
AdS-wave log solutions may provide a dual LCFT$_3$, while at
off-critical point,  its dual theory may correspond to a
nonrelativistic field theory with fixed boundary conditions.  It
suggests  that  a similar thing may happen at the tricritical point
and off-tricritical point in the 4DGMG.   However, an explicit
construction of its dual CFT  is beyond the scope of the present
work and thus, it is left as a future work.

 \vspace{1cm}

{\bf Acknowledgments}

This work was supported by the National Research Foundation of Korea
(NRF) grant funded by the Korea government (MEST) through the Center
for Quantum Spacetime (CQUeST) of Sogang University with grant
number 2005-0049409. Y. Myung  was partly  supported by the National
Research Foundation of Korea (NRF) grant funded by the Korea
government (MEST) (No.2012-040499).


\newpage
\begin{thebibliography}{99}

\bibitem{Stelle}
  K.~S.~Stelle,
  Phys.\ Rev.\  D {\bf 16}, 953 (1977).

\bibitem{Li:2008dq}
  W.~Li, W.~Song and A.~Strominger,
   JHEP {\bf 0804}, 082 (2008)  [arXiv:0801.4566 [hep-th]].

\bibitem{Lu:2011zk}
  H.~Lu and C.~N.~Pope,
  Phys.\ Rev.\ Lett.\  {\bf 106}, 181302 (2011)  [arXiv:1101.1971 [hep-th]].

\bibitem{Deser:2011xc}
  S.~Deser, H.~Liu, H.~Lu, C.~N.~Pope, T.~C.~Sisman and B.~Tekin,
  Phys.\ Rev.\ D {\bf 83}, 061502 (2011)  [arXiv:1101.4009 [hep-th]].

\bibitem{Porrati:2011ku}
  M.~Porrati and M.~M.~Roberts,
   Phys.\ Rev.\ D {\bf 84}, 024013 (2011)  [arXiv:1104.0674 [hep-th]].

\bibitem{Alishahiha:2011yb}
  M.~Alishahiha and R.~Fareghbal,
  Phys.\ Rev.\ D {\bf 83}, 084052 (2011)  [arXiv:1101.5891 [hep-th]].

\bibitem{Bergshoeff:2011ri}
  E.~A.~Bergshoeff, O.~Hohm, J.~Rosseel and P.~K.~Townsend,
  Phys.\ Rev.\ D {\bf 83}, 104038 (2011)  [arXiv:1102.4091 [hep-th]].


\bibitem{Lu:2011mw}
  H.~Lu, C.~N.~Pope, E.~Sezgin and L.~Wulff,
  JHEP {\bf 1110}, 131 (2011)  [arXiv:1107.2480 [hep-th]].


\bibitem{Grumiller:2008qz}
  D.~Grumiller and N.~Johansson,
   JHEP {\bf 0807}, 134 (2008)  [arXiv:0805.2610 [hep-th]].
\bibitem{Myung:2008dm}
  Y.~S.~Myung,
  Phys.\ Lett.\ B {\bf 670}, 220 (2008)  [arXiv:0808.1942 [hep-th]].

\bibitem{Maloney:2009ck}
  A.~Maloney, W.~Song and A.~Strominger,
  Phys.\ Rev.\ D {\bf 81}, 064007 (2010)  [arXiv:0903.4573 [hep-th]].

\bibitem{Gurarie:1993xq}
  V.~Gurarie,
    Nucl.\ Phys.\ B {\bf 410}, 535 (1993)  [hep-th/9303160].

\bibitem{Flohr:2001zs}
  M.~Flohr,
   Int.\ J.\ Mod.\ Phys.\ A {\bf 18}, 4497 (2003)  [hep-th/0111228].


\bibitem{Nutma:2012ss}
  T.~Nutma,
  Phys.\ Rev.\ D {\bf 85}, 124040 (2012)  [arXiv:1203.5338 [hep-th]].

\bibitem{Bergshoeff:2012sc}
  E.~A.~Bergshoeff, S.~de Haan, W.~Merbis, M.~Porrati and J.~Rosseel,
   JHEP {\bf 1204}, 134 (2012)  [arXiv:1201.0449 [hep-th]].


\bibitem{Bergshoeff:2012ev}
  E.~A.~Bergshoeff, S.~de Haan, W.~Merbis, J.~Rosseel and T.~Zojer,
   arXiv:1206.3089 [hep-th].

\bibitem{Johansson:2012fs}
  N.~Johansson, A.~Naseh and T.~Zojer,
  arXiv:1205.5804 [hep-th].

\bibitem{Kleinschmidt:2012rs}
  A.~Kleinschmidt, T.~Nutma and A.~Virmani,
  arXiv:1206.7095 [hep-th].

\bibitem{Apolo:2012vv}
  L.~Apolo and M.~Porrati,
   arXiv:1206.5231 [hep-th].

\bibitem{LPP}
  H.~Lu, Y.~Pang and C.~N.~Pope,
   Phys.\ Rev.\ D {\bf 84}, 064001 (2011)  [arXiv:1106.4657 [hep-th]].


\bibitem{LS}
  Y.~Liu and Y.~-W.~Sun,
  Phys.\ Rev.\ D {\bf 79}, 126001 (2009)  [arXiv:0904.0403 [hep-th]].

\bibitem{Deser:1981wh}
  S.~Deser, R.~Jackiw and S.~Templeton,
   Ann. Phys. (N.Y.)\  {\bf 140}, 372 (1982); {\bf 185}, 406(E)
   (1988); {\bf 281}, 409 (2000).

\bibitem{Bergshoeff:2009hq}
  E.~A.~Bergshoeff, O.~Hohm and P.~K.~Townsend,
   Phys.\ Rev.\ Lett.\  {\bf 102}, 201301 (2009)  [arXiv:0901.1766 [hep-th]].


\bibitem{Grumiller:2010tj}
  D.~Grumiller, N.~Johansson and T.~Zojer,
   JHEP {\bf 1101}, 090 (2011)  [arXiv:1010.4449 [hep-th]].

\bibitem{AyonBeato:2009yq}
  E.~Ayon-Beato, G.~Giribet and M.~Hassaine,
   JHEP {\bf 0905}, 029 (2009)  [arXiv:0904.0668 [hep-th]].

\bibitem{KMP}
  Y.~-W.~Kim, Y.~S.~Myung and Y.~-J.~Park,
    arXiv:1207.3149 [hep-th].


\bibitem{Jackiw:2003pm}
  R.~Jackiw and S.~Y.~Pi,
  Phys.\ Rev.\ D {\bf 68}, 104012 (2003)
  [gr-qc/0308071].


\bibitem{Alexander:2009tp}
  S.~Alexander and N.~Yunes,
  Phys.\ Rept.\  {\bf 480}, 1 (2009)  [arXiv:0907.2562 [hep-th]].


\bibitem{Cardoso:2009pk}
  V.~Cardoso and L.~Gualtieri,
  Phys.\ Rev.\ D {\bf 80}, 064008 (2009)
  [Erratum-ibid.\ D {\bf 81}, 089903 (2010)]
  [arXiv:0907.5008 [gr-qc]].


\bibitem{Moon:2011fw}
  T.~Moon and Y.~S.~Myung,
  Phys.\ Rev.\ D {\bf 84}, 104029 (2011)
  [arXiv:1109.2719 [gr-qc]],
and references therein.


\bibitem{MM}
  T.~Moon and Y.~S.~Myung,
  Eur.\ Phys.\ J.\ C {\bf 71}, 1796 (2011)
  [arXiv:1108.2612 [hep-th]].


\bibitem{beato}
 E. A.-Beato, G. Giribet, and M. Hassaine,
 Phys. Rev. {\bf D83}, 104033 (2011)
 [arXiv:1103.0742 [hep-th]].


\bibitem{Gullu:2011sj}
  I.~Gullu, M.~Gurses, T.~C.~Sisman and B.~Tekin,
  Phys.\ Rev.\ D {\bf 83}, 084015 (2011)
  [arXiv:1102.1921 [hep-th]].


\bibitem{Tekin:2007rn}
  B.~Tekin,
  Phys.\ Rev.\ D {\bf 77}, 024005 (2008)  [arXiv:0710.2528 [gr-qc]].



\bibitem{AyonBeato:2005qq}
  E.~Ayon-Beato and M.~Hassaine,
  Phys.\ Rev.\ D {\bf 73}, 104001 (2006)  [hep-th/0512074].


\end{thebibliography}
\end{document}